\documentclass[aps,prl,twocolumn,groupedaddress]{revtex4}

\usepackage {graphicx}

\begin{document}


\title{
Insulating Ferromagnetism in La$_{4}$Ba$_{2}$Cu$_{2}$O$_{10}$:\\
an \textit{Ab Initio} Wannier Function Analysis
}


\author{Wei Ku}

\author{H. Rosner}
\author{W.E. Pickett}
\author{R.T. Scalettar}
\affiliation{
Department of Physics, University of California, Davis, CA 95616-8677
}



\begin{abstract}
Microscopic mechanisms of the puzzling insulating ferromagnetism of half-filled
La$_{4}$Ba$_{2}$Cu$_{2}$O$_{10}$ are elucidated with energy-resolved Wannier states.
The dominant magnetic coupling, revealed through evaluated parameters ($t$, $U$, and $J$), 
turns out to be the intersite direct exchange, a currently ignored mechanism
that overwhelms the antiferromagnetic superexchange.
By contrast, the isostructural Nd$_{4}$Ba$_{2}$Cu$_{2}$O$_{10}$ develops the observed 
antiferromagnetic order via its characteristics of a 1D chain.
Surprisingly, the in-plane order of both cases is \textit{not} controlled by 
coupling between nearest neighbors.
An intriguing pressure-induced ferromagnetic to antiferromagnetic transition is predicted.
\end{abstract}


\maketitle

\begin{figure}[!tbp]
\includegraphics[width=3.00in]{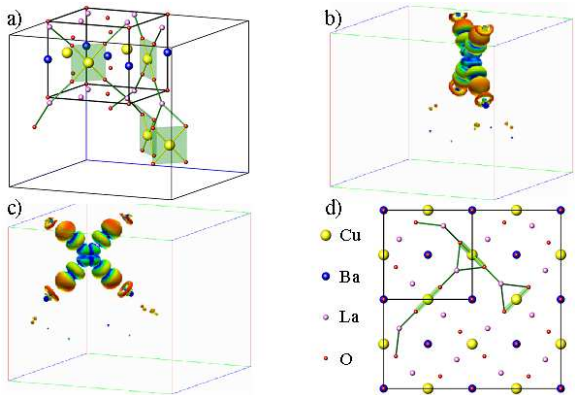}
\caption{
\label{fig1}
Crystal structure of La422 (a) and its top view (d)
shown within $2\times 2\times2$ unit cells.
Most atoms outside the 1$^{st}$ unit cell are removed from panel a) for clarity.
b) and c): illustration of low-energy Wannier states with $R = \left( {0,0,0} \right)$.
}
\vspace*{-0.5cm}
\end{figure}

Unlike most half-filled cuprates that feature the antiferromagnetic (AF) 
ground state as expected from the Hubbard model,
few of the cuprates possess exceptional ferromagnetic (FM) order instead \cite{Mizuno,Pieper,deJongh}.
In particular, half-filled Cu-$d$ orbitals in the ``brown phase'' La$_{4}$Ba$_{2}$Cu$_{2}$O$_{10}$
(La422) is found to align ferromagnetically below 5K \cite{Mizuno,Pieper},
in great contrast to the isostructural (Fig.~\ref{fig1}) Nd$_{4}$Ba$_{2}$Cu$_{2}$O$_{10}$
(Nd422) that retains antiferromagnetism below 7.5K \cite{Paukov,Nozaki,Golosovsky}.
Such an unusual insulating FM phase poses a great challenge to our \textit{quantitative}
theoretical understanding of microscopic mechanisms involved in real materials.

Up to now, there are only limited pioneering attempts at identifying the quantum processes responsible
for this intriguing behavior of La422.
With elaborate perturbation on a \textit{multiband} Hubbard-like model \cite{Feldkemper},
it is suggested that destructive interference of hopping paths (between 6$^{th}$ order and 8$^{th}$
order terms) may suppress AF coupling and give a small FM coupling along the $z$-axis
in a very narrow parameter range, while the Goodenough process \cite{Goodenough} between nearest
neighbors produces the in-plane FM order.
(More recently, the possible ``crude link'' to the ``flat-band ferromagnetism'' from the Hubbard model
was also pointed out \cite{Tasaki}.)
However, this description is not very satisfactory;
besides the problem of the narrow parameter range,
the resulting strength of the FM coupling is too small to account for the experiments \cite{Mizuno}.
Furthermore, the assumption of FM coupling between the nearest neighbors in the plane is apparently
in contradiction to the AF order in Nd422 \cite{Golosovsky}.

From the theoretical point of view, descriptions of quantum magnetism have been given mainly via
phenomenological models like the Stoner, Heisenberg, Hubbard, or other
tight-binding models \cite{Goodenough,Hirsch}, with adjustable parameters to fit the experiments.
These approaches, though intuitive and computationally manageable, lack the quantitative detail of
the complex interactions that occur in real materials.
Sometimes even the sign of spin-spin coupling is simply chosen to fit the experimental magnetic order.
On the other hand, a normal practice of more quantitative \textit{ab initio} approaches \cite{Eyert},
involving comparing total energy corresponding to different magnetic orders within the density
functional theory (DFT) \cite{Kohn},
not only suffers from the uncertain quality of existing approximate energy functionals \cite{Kohn} for
the very sensitive magnetic systems, but also hides all the microscopic processes in a black box.
The development of a unified scheme that simultaneously gives intuitive microscopic insight and
quantitative realism of \textit{ab initio} level is thus highly desired.

In this letter, aiming to resolve specifically the microscopic mechanisms of the unusual insulating
ferromagnetism in La422,
we attempt the first step of one such ambitious scheme by evaluating parameters ($t$, $U$, and $J$)
of (reformulated) \textit{ab initio} second quantized Hamiltonian in the basis of localized
energy-resolved Wannier States (WSs) \cite{Wannier,Marzari,Smirnov,Berghold,McMahan},
constructed from all-electron DFT orbitals \cite{Kohn}.
A clear picture of microscopic processes involved naturally emerges from 
these parameters and the spatial distribution of the WSs.
The dominant mechanism turns out to be the intersite FM ``direct exchange'' \cite{Hirsch}
that is currently ignored in the microscopic studies of this system.
This process, occurring mainly at the La and O sites, overwhelms the weak tendency toward AF order
via superexchange of Hubbard type \cite{Anderson}.
The same analysis is then applied to Nd422 and used to demonstrate its characteristics of 1D AF chain,
as experimentally observed \cite{Paukov,Golosovsky}.
Surprisingly, in both compounds, the dominant (FM) exchange is found \textit{not} with the nearest 
neighbors (currently assumed), but with sites above/below them,
which simultaneously generates the proper in-plane order.
Finally, the crucial role of ``chemical'' effect is illustrated with a numerical simulation of La422
``under pressure'', which suggests an intriguing pressure-induced FM to AF transition.
Predictions are made for further experimental investigation.

In order to uncover the underlying many-body interactions responsible for the observed magnetic
order, mapping the system onto a simplified effective model Hamiltonian \cite{McMahan,Hybertsen}
is avoided.
Instead, attention is paid to the full \textit{ab initio} Hamiltonian, \textit{rigorously}
reformulated into a ``fluctuation'' form that explicitly utilizes DFT solutions \cite{Ku}
(without ``double-counting'' many-body effects):
\begin{eqnarray}
&H = \varepsilon _{\bar {\mu }\bar {\alpha }} n_{\bar {\mu }\bar {\alpha }} 
- t_{\bar {\mu }\bar {\nu }} c_{\bar {\mu }\bar {\alpha }}^ + c_{\bar {\nu 
}\bar {\alpha }}& \nonumber\\
 &+ U_{\bar {\mu }} \tilde {n}_{\bar {\mu } \uparrow } \tilde {n}_{\bar {\mu } 
\downarrow } + C_{\bar {\mu }\bar {\nu }} \tilde {n}_{\bar {\mu }\bar 
{\alpha }} \tilde {n}_{\bar {\nu }\bar {\beta }} - J_{\bar {\mu }\bar {\nu 
}} \left( {S_{\bar {\mu }} \cdot S_{\bar {\nu }} + \frac{1}{4}n_{\bar {\mu }\bar 
{\alpha }} n_{\bar {\nu }\bar {\beta }} } \right)& \nonumber\\
 &+ \mbox{other }\widetilde{ \left( c_{\bar {\mu }\bar {\alpha }}^ + c_{{\bar {\mu 
}}'\bar {\alpha }} \right) }\widetilde{ \left( c_{\bar {\nu }\bar {\beta }}^ + c_{{\bar {\nu 
}}'\bar {\beta }} \right) }\mbox{ terms} + \mbox{constant terms,}\quad &
\label{eq1}
\end{eqnarray}
where summation over barred variables is understood and
$\tilde{o}\equiv o-\left\langle o\right\rangle^{DFT}$ denotes ``fluctuation'' from expectation
value $\left\langle o \right\rangle ^{DFT}$ of operator $o$ within DFT.
Notice that the direct exchange (the $5^{th}$ term) that directly couples spins is omitted in all
the existing Hubbard-model based analysis of La422, and turns out to play a crucial role.

Extraction of parameters of this full Hamiltonian via traditional constraint DFT calculations
\cite{McMahan,Hybertsen} is extremely complicated (if even possible.)
Instead, ``bare'' parameters
$t_{\mu\nu}\equiv -\left\langle{\mu\left|{h^{DFT}-v^{xc}}\right|\nu}\right\rangle
\left({1-\delta_{\mu\nu}}\right)
+ J_{\mu\nu}\langle{c_{\nu\bar{\beta}}^+c_{\mu\bar{\beta}}}\rangle^{DFT}
\sim -\left\langle{\mu\left|{h^{DFT}}\right|\nu}\right\rangle \left({1-\delta_{\mu\nu}}\right)$,
$J_{\mu\nu}\equiv\left\langle{\mu\nu\left|v\right|\nu\mu}\right\rangle\left({1-\delta_{\mu\nu}}\right)$,
and $U_{\mu}\equiv\left\langle{\mu\mu\left|v\right|\mu\mu}\right\rangle$
are \textit{directly evaluated} on the basis of half-filled, low-energy, localized WSs that possess
the spin moment.
($h^{DFT}$, $v^{xc}$, and $v$ are the DFT Hamiltonian, exchange-correlation potential, and bare Coulomb
interaction.)
These parameters prove to be insightful to illustrate the insulating ferromagnetism of interest, 
as they control how electrons (and spins) interact with each other (in a many-body environment)
in real materials.
Now, since solving the multiband Eq.~(\ref{eq1}) currently remains an important and yet
challenging task, we proceed by identifying two primary microscopic mechanisms that appear naturally
in Eq.~(\ref{eq1}): Hubbard-type AF superexchange \cite{Anderson} and FM direct exchange \cite{Hirsch}
(between the WSs).
Their strength can be estimated by summing \textit{phenomenological} formulae:
${\mathcal J}^{SX} \sim - 4t_{ij}^2 / W_{j}$ and ${\mathcal J}^{DX} \sim J_{ij} / 2$,
respectively, over leading exchange paths, where $i$ and $j$ are site index and $W_{j} = U_{j} / 2$
approximately accounts for effects of on-site repulsion, competing off-site Coulomb repulsion,
and other possible virtual processes involving states of higher energy (screening.)
(The factor in $W_{j}$ is chosen to give roughly consistent size of the expected Mott-Hubbard gap in
cuprates.
Considering the \textit{intersite} nature of the direct exchange, the factor in ${\mathcal J}^{DX}$ is
reasonable \cite{Marel}.
More accurate estimation of these factors will not qualitatively change our conclusion on the
insulating ferromagnetism of La422.)

The employment of the WSs (or ``molecular orbitals in crystal'')
\cite{Wannier,Marzari,Smirnov,Berghold,McMahan} as basis,
instead of other orthonormal complete sets, is strongly physically motivated.
In addition to full awareness of crystal and local symmetries \cite{Smirnov}, the constructed WSs are
\textit{the most localized within the subspace of low-energy excitation}, which facilitate direct
physical interpretation consistent with localized spins interacting with each other.
This ``energy resolution'' naturally follows the identity of the subspace spanned by our WSs and
that by the \textit{chosen} eigenstates of $h^{DFT}$.
In addition, single-particle properties of these WSs
---(``self-interaction''\cite{Perdew} free) orbital energy
$\varepsilon_{\mu\alpha}\equiv\left\langle{\mu\left|{h^{DFT}-v^{xc}}\right|\mu}\right\rangle
 - U_\mu\left\langle{n_{\mu\alpha}}\right\rangle^{DFT}$,
occupation number, and spin moment--- are directly translated with information from the DFT calculation,
and are known \textit{before} solving the many-body Hamiltonian.
Specifically for La422, we are thus able to focus only on the \textit{half-filled} low-energy WSs
corresponding to the \textit{two} partially occupied bands across the Fermi energy,
as other particle/hole states are of much higher energy and with zero spin moment.
Furthermore, superexchange involving other states diminishes due to orthogonality between subspaces.

In this work, $h^{DFT}$ is chosen via LDA \cite{Perdew92} for La422 and via LSDA+U \cite{Anisimov} for
Nd422 (to prevent Nd $f$-states from incorrectly falling on the Fermi energy.)
Values of $U$ and $J$ (chosen to be 6.7 and 0.7 eV) in LSDA+U, employed to split the
$f$-states, are not crucial as long as their hybridization with the WSs is eliminated.

The WSs in unit cell $R$ and of orbital $m$ are constructed by $\left|{Rm}\right\rangle
\equiv\left|{\bar{k}m}\right\rangle e^{-i\bar{k}\cdot R}\mbox{/}\sqrt N$,
where $N$ is the number of discrete $k$-points in the reciprocal space,
and $\left|{km}\right\rangle=\left|{\psi_{k{\bar{m}}'}}\right\rangle
\left\langle{\psi_{k{\bar{m}}'}}|km\right\rangle$
(band-mixed Bloch state of crystal momentum $k)$ is transformed from the
all-electron \cite{PSApprx} DFT eigenstate $\left|{\psi_{km'}}\right\rangle$ \cite{Blaha} of band index
${m}'$, to facilitate maximum localization of the WSs.
We follow the first step in Ref.~\cite{Marzari} to calculate $\left\langle{\psi_{km'}}|km\right\rangle$
by projecting onto the \textit{chosen subspace} of $\left|{\psi_{km'}}\right\rangle$
(belonging to the two partially occupied bands) narrow Gaussian states,
$\left|{g_m}\right\rangle$, each centered at one of the main lobes of Cu-$d$ contribution,
followed by symmetric orthogonalizaton: $\left\langle\psi_{km'}|km\right\rangle =
\left\langle\psi_{km'}|g_{{\bar{m}}''}\right\rangle M_{{\bar{m}}''m}$;
$M_{m''m}^{-2} \equiv \left\langle{g_{m''}}|\psi_{k{\bar{m}}'}\right\rangle
\left\langle\psi_{k{\bar{m}}'}|g_m\right\rangle$.
Due to the concentrated contribution from Cu-$d$ orbitals and the fact that there is only one
low-energy WS per Cu site, the resulting WSs are practically the same as the maximally
localized WSs constructed with the complete iteration procedure \cite{Marzari}.
While denser $k$-sampling can slightly refine the shape of the resulting WSs, a $7\times 7\times8$ 
lattice is good enough to give reliable description, as the ``aliasing'' caused 
by overlap of WS with itself due to periodicity is eliminated.

The resulting low-energy WSs of La422, each centered at one Cu site, are shown in Fig.~\ref{fig1},
in which isosurfaces of $\left|{\langle x|Rm\rangle}\right|^2$ in real space $x$ are plotted,
along with the crystal structure.
Notice that the ``hybridization'' between Cu-$d$, O-$p$ and La-$d$ orbitals is naturally built into
each WS; in the traditional atomic-orbital based analysis \cite{McMahan}, such hybridization would
require inclusion of complicated hopping and interaction between several orbitals from Cu, O, and La
covering larger energy, which in turn obscures straightforward visualization of key physical
processes.
The anti-bonding nature of the Cu-$d$ and O-$p$ orbitals is also easily observed, as well as the 
fascinating ``ring'' structure at the La sites, which turns out to play an important role.

The spatial distribution of the WSs also reflects that of the spin density:
$\left\langle{x|\bar{R}\bar{m}}\right\rangle
\left\langle{c_{\bar{R}\bar{m}\alpha}^+ c_{{\bar{R}}'{\bar{m}}'\alpha}}\right\rangle
\left\langle{{\bar{R}}'{\bar{m}}'|x}\right\rangle \sim
\left|{\langle x|\bar{R}\bar{m}\rangle}\right|^2 \left\langle{n_{\bar{R}\bar{m}\alpha}}\right\rangle$
for localized WSs.
We found that about 50{\%} of the spin-moment resides at the Cu site, while 10{\%} at each
of the four O near Cu and 2{\%} ($\times 2$ from two WSs) at each La;
the last naturally accounts for the experimentally observed hyperfine field for La \cite{Pieper}.
Our WSs also illustrate, for the first time, the shape of the spin density at the La sites
(as superposition of $d$ orbitals of various symmetry),
which has not been completely resolved by NMR measurement \cite{Pieper}.

\begin{figure}[!bp]
\vspace*{-0.5cm}
\includegraphics[width=3.50in]{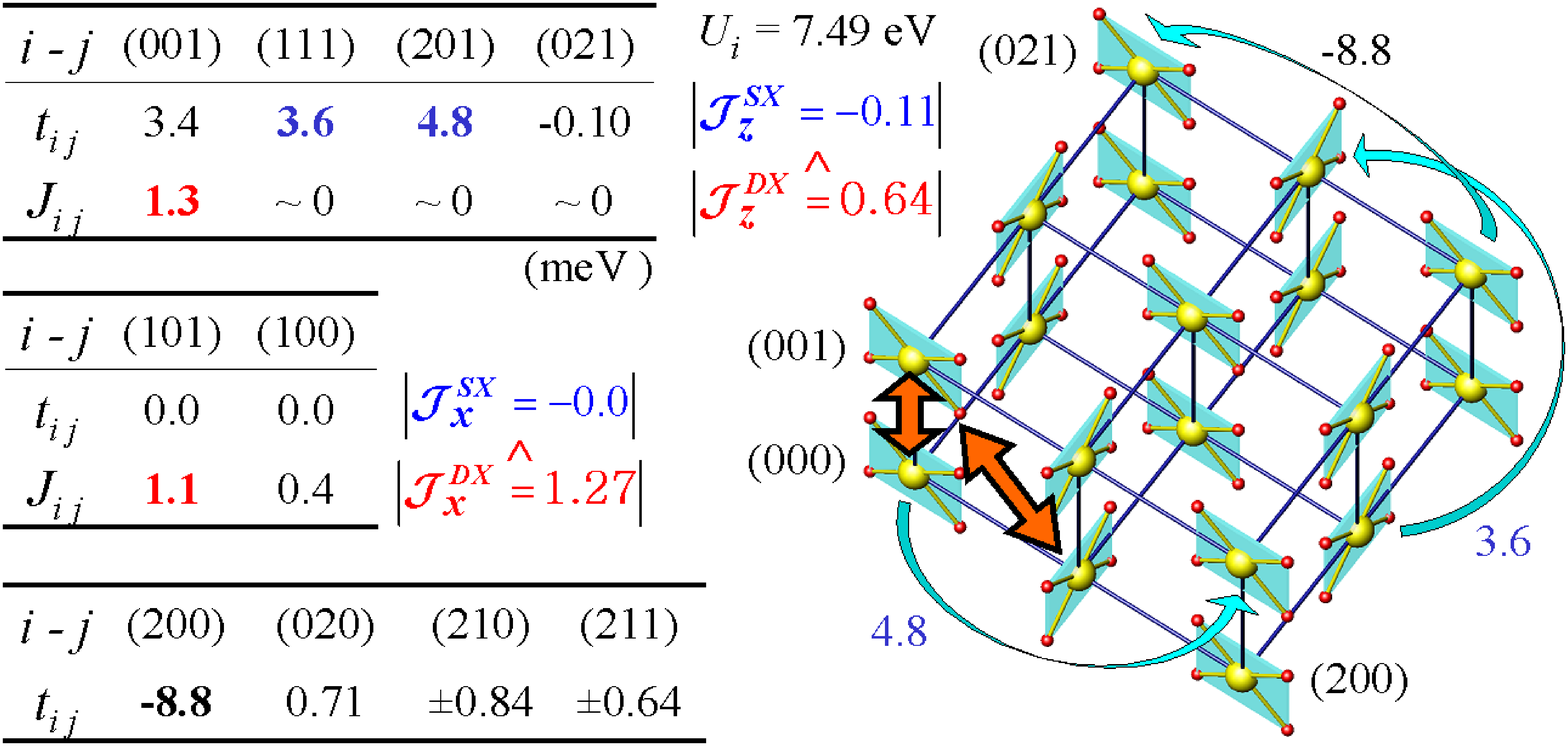}
\caption{
\label{fig2}
Calculated parameters (in meV) 
for La422 and schematics of
leading hopping (blue single arrows) and direct exchange (red double arrows) channels.
Coordinate system for site indices $i$ and $j$ are defined for convenience.
}
\vspace*{-0.5cm}
\end{figure}

Figure~\ref{fig2} lists our calculated leading parameters.
Note that $t_{Rm,{R}'{m}'}\sim\left\langle{Rm|\bar{k}{\bar{m}}''}\right\rangle
\varepsilon_{\bar{k}{\bar{m}}''} \left\langle{\bar{k}{\bar{m}}''|{R}'{m}'}\right\rangle$
is easily \textit{evaluated} with DFT eigenenergy $\varepsilon_{k{m}''}$, in contrast to the
conventional procedure of tight-binding \textit{fit} \cite{Papa}.
(Since the resulting $t_{ij}$ corresponds \textit{exactly} to the original DFT eigenenergy,
this can be viewed as perfect "downfolding" \cite{Andersen}.)
$U_i$, and $J_{ij}$ are obtained via careful numerical integration \cite{Ku,IntApprx},
instead of parameterized Slater integrals \cite{Slater} that assume \textit{atomic} orbitals
with well-defined angular momentum.
The resulting $U_i \sim 7.5$ eV is much smaller than that obtained from atomic Cu-$d$ orbitals,
reflecting non-negligible hybridization.

The microscopic mechanisms for the unusual FM order are now easily visualized.
Consistent with the extremely narrow bandwidth of the low energy bands, hopping ($t_{ij} )$ between
the WSs is very small (Fig.~\ref{fig2}), of a few meV,
which brings about a weak AF tendency via superexchange \cite{Anderson}.
Along $z$-axis (1$^{st}$ table), the FM direct exchange, commonly discarded in Hubbard-like models,
clearly overwhelms the AF superexchange (${\mathcal J}_z^{DX} \gg {\mathcal J}_z^{SX}$).
This process occurs mainly at the ``ring'' structure of WSs at the La sites (Fig.~\ref{fig1}),
where a large hyperfine field occurs \cite{Pieper}.
In the language of atomic physics, this direct exchange is similar to what gives the first
Hund's rule at La sites.

Interestingly, instead of the nearest neighbor interaction currently assumed by existing studies,
the in-plane FM coupling (2$^{nd}$ table) turns out to be dominated
by the (101) direct exchange (\textit{above}/\textit{below} the nearest neighbors),
occurring mainly at O sites near Cu (Fig.~\ref{fig1}.)
Having the interplane FM alignment, a strong in-plane FM coupling is thus established,
unchallenged by any AF coupling, due to zero hopping dictated by the local symmetry \cite{Smirnov}.
The resulting coupling constants are of the same order as the observed transition
temperature ($\sim$5K), which further rationalizes our phenomenological estimation.

\begin{figure}[!bp]
\vspace*{-0.5cm}
\includegraphics[width=3.50in]{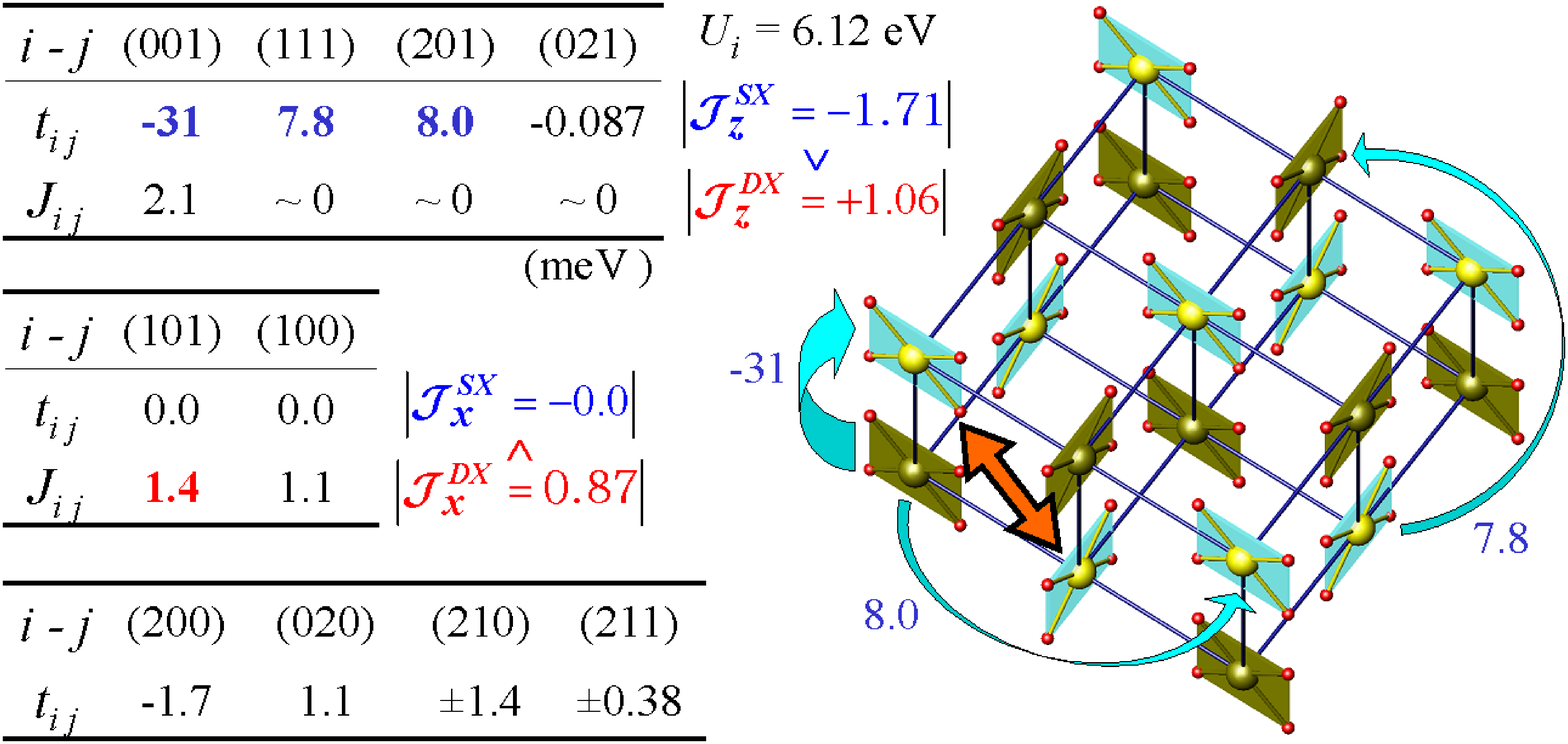}
\includegraphics[width=3.50in]{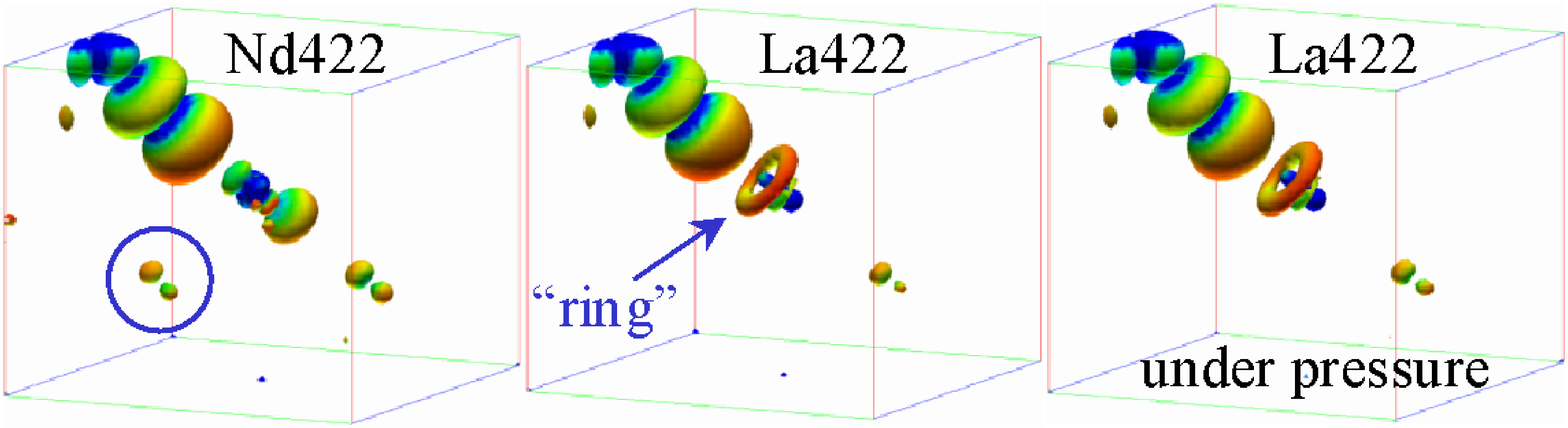}
\caption{
\label{fig3}
Upper panel: calculated parameters (in meV) with low-energy WSs of Nd422
(c.f.: Fig.~\ref{fig2}).
Lower panel: comparison of WSs of Nd422, La422, and La422 under pressure.
}
\vspace*{-0.5cm}
\end{figure}

For comparison, the AF order in Nd422 is also analyzed with the same technique.
(According to the analysis of neutron diffraction measurement \cite{Golosovsky}, the Nd moment is
perpendicular to and ``disconnected'' from Cu moment, and thus is excluded in our analysis.)
One immediately notices in Fig.~\ref{fig3} that the (001) hopping is considerably ($\sim $10 times)
larger, which in turn greatly enhances the AF  superexchange and thus allows it to overcome the
(slightly enhanced) FM direct coupling along the $z$-axis.
Such strong 1D characteristics is in excellent agreement with the optical and neutron measurement
\cite{Paukov,Golosovsky}.
In the $xy$-plane, however, there exits \textit{no} AF coupling between nearest neighbors
but only FM exchange, primarily along (101), as in the case of La422.
(The competing (100) FM coupling is weaker, and connects to fewer neighbors.)
We thus reach a surprising conclusion that, having the AF order between planes,
\textit{the observed in-plane AF order is actually caused by the (101) FM coupling}.

Insight into the dramatic difference in the $z$-axis hopping in these two compounds can be
obtained by comparing the corresponding WSs, shown in Fig.~\ref{fig3}.
In Nd422, the development of extra $p$-contribution (circled) at O sites that belong to the WS
one layer below/above, provides additional efficient hopping channels along the $z$-axis.
By contrast, the weak $z$-axis hopping in La422 is mainly through the (inefficient) ``ring'' structure
at the La site, and is further weakened by the residual hopping via O sites (not shown in
Fig.~\ref{fig3}), due to the phase difference between these two different routes.
(This difference in hopping channels is clearly reflected in the sign of (001) hopping,
since $t_{ij} $, unlike $J_{ij} $, is sensitive to the phase shift.)
This weakening effect resembles the ``interference'' discussed in Ref. \cite{Feldkemper},
except that in our analysis it is not responsible for the observed FM order.

The driving mechanism of this spatial redistribution of charge/spin is numerically
examined by another set of calculations for La422 with lattice constants and
atomic positions identical to Nd422 (La422 ``under pressure'' with $\sim$ 2{\%} reduction of the
lattice constants.)
The resulting WSs, shown in Fig.~\ref{fig3}, do not possess the extra O-$p$ structure.
Consistently, the key $z$-axis hopping only reaches 30{\%} of that in Nd422 and the sign of (001)
hopping remains positive.
Evidently, the charge redistribution is mainly driven by the change of chemical environment
surrounding Nd, which is smaller in size and consists of 3 $f$-electrons and 3 more protons.

Intriguingly, while the resulting magnetic order remains FM, in good agreement with experiment on La422
under pressure \cite{Nozaki}, moderate enhancement of $t_{ij}$ and $J_{ij}$ is observed due to
slightly larger covalency, i.e. overlap between WSs.
As a result, AF superexchange starts to approach the strength of the FM direct exchange (since $t$ is
squared) indicating a FM to AF transition upon \textit{higher} pressure.
We propose further experiment with pressure slightly higher than 9GPa to observe this fascinating
FM to AF transition.
Alternatively, this transition should be more easily achievable with moderate pressure on
(La$_{0.8}$Nd$_{0.2}$)$_4$Ba$_2$Cu$_2$O$_{10}$.

In summary, microscopic mechanisms of the puzzling insulating FM order observed in La422 are identified
based on calculated parameters of interactions between half-filled localized WSs.
The long-omitted FM direct exchange is shown to overwhelm AF superexchange.
The spatial distribution of the constructed low-energy all-electron WSs provides detailed insight into
the exchange processes and the partial contribution of spin-moment, in agreement with the
NMR measurement.
The isostructural Nd422 is shown to possess characteristics of 1D AF chain along $z$-axis,
as experimentally observed, via extra efficient hopping through O sites.
The in-plane magnetic order of both compounds turns out to be introduced via dominant FM coupling
with (101) neighbors, not the nearest ones currently assumed.
The crucial role of chemical replacement is illustrated via numerical test on La422 ``under pressure'',
which further predicts an intriguing pressure-induced FM to AF transition.
While future development for the many-body solution is desired, our scheme already
allows detailed, quantitative, and intuitive description of quantum magnetism
that complements the common model approaches.

\begin{acknowledgments}
This work was supported by DOE Grant DE-FG03-01ER45876, ASCI through LLNL, and DAAD.
Critical comments from R.R.P. Singh and discussion with A.K. McMahan and D.D. Koelling are acknowledged.
\vspace*{-0.8cm}
\end{acknowledgments}


\newpage
\printtables
\newpage
\printfigures

\end{document}